\documentstyle[psfig]{aa}
\begin{document}

\title{M31-RV evolution and its alleged multi-outburst 
        pattern\thanks{Table~3 available only in electronic form (ASCII format) 
        at CDS via anonymous ftp to cdsarc.u-strasbg.fr (130.79.128.5) or via
        http://cdsweb.u-strasbg.fr/cgi-bin/qcat?J/A+A/ and from the web page
        http://ulisse.pd.astro.it/M31-RV/, where further information is provided}
       }

\author{
       Federico Boschi 
	\and
       Ulisse Munari
       }
\offprints{U.Munari}

\institute {
INAF Osservatorio Astronomico di Padova, Sede di Asiago, I-36012 Asiago (VI), Italy
           }

\date{Received date..............; accepted date................}

\abstract{
The photometric evolution of M31-RV has been investigated on 1447 plates of
the Andromeda galaxy obtained over half a century with the Asiago
telescopes. M31-RV is a gigantic stellar explosion that occurred during 1988
in the Bulge of M31 and that was characterized by the appearance for a few
months of an M supergiant reaching $M_{bol}=-$10. The 1988 outburst has been
positively detected on Asiago plates, and it has been the only such event
recorded over the period covered by the plates (1942-1993). In particular, an alleged
previous outburst in 1967 is excluded by the more numerous and deeper Asiago
plates, with relevant implication for the interpretative models of this
unique event. We outline a close analogy in spectral and photometric
evolution with those of V838~Mon which exploded in our Galaxy in 2002. The
analogy is found to extend also to the closely similar absolute magnitude at
the time of the sudden drop in photospheric temperature that both M31-RV and
V838~Mon exhibited. These similarities, in spite of the greatly differing
metallicity, age and mass of the two objects, suggest that the same, universal
and not yet identified process was at work in both cases.

         }

\maketitle

\section{Introduction}

Rich et al. (1989) discovered in 1988 a highly unusual stellar outburst in
the Bulge of the Andromeda galaxy (M31), known since then as M31-RV (for
``red variable''). The event peaked at M$_{bol} \approx -10$~mag and
its spectrum closely resembled that of M supergiants, evolving from M0~I at
discovery (Sept 5, 1988) to $>$M7~I about 58 days later when the brightness
in the $V$ band had dropped by at least 4 mag (Rich 1990). Two similar
events have been later identified in our Galaxy, V4332~Sgr that exploded in
1994 (Martini et al. 1999) and V838~Mon that erupted in 2002 (Munari et al.
2002a, Bond et al. 2003, and references therein).

The M31-RV event has been characterized by radiative luminosities in-between
those of classical novae and supernovae. The mass of the ejected envelope
(optically thick during the whole observed evolution) is uncertain but it is
certainly larger than in typical novae and much less than in supernovae. The
radiative and kinetic energetics place therefore M31-RV, and by analogy also
V4332~Sgr and V838~Mon, in the gap between classical novae and supernovae,
making them stars of special interest. So far, few theoretical attempts to
explain their highly peculiar nature have been pursued. Soker and Tylenda
(2003), to explain the energetics and multi-maxima behaviour of V838~Mon,
have suggested the merging of two main sequence stars of masses
0.1$-$0.5~M$_\odot$ and 1.5~M$_\odot$, with the second one expanding to
large radii, low temperature and high luminosity in response to the
frictional energy dissipation of the cannibalized less massive companion. A
similar scenario has been proposed by Retter and Marom (2003). They
postulated the multi-maximum eruption of V838~Mon as the result of the
swallowing of massive planets in close orbit around a parent star expanding
while on the RGB (red giant branch) or AGB (asymptotic giant
branch). A thermonuclear runaway (TNR) model was instead developed by Iben and
Tutukov (1992) to explain M31-RV. The model envisages a binary system,
composed of a WD and a low mass companion, that evolves to orbital periods
shorter than 2 hours by loss of angular momentum via gravitational waves,
without experiencing classical nova eruptions on the way to. The accretion
at very low rates ($\sim 10^{-11}$~M$_\odot$~yr$^{-1}$) occurring onto a
cold white dwarf (WD) can lead to the accumulation of a massive H-rich envelope of the
order of $\sim$0.05~M$_\odot$ before this is expelled in a gigantic hydrogen
shell flash (some 10$^3$ times the mass expelled in a typical nova
eruption). Friction energy dissipation of the binary revolving within such a
massive and dense common envelope can raise the drag luminosity to
10$^7$~L$_\odot$, with as much as 10$^6$~L$_\odot$ ($\gg$~L$_{\rm
Eddington}$) coming out in the form of radiation.
 
     \begin{table*}[!t]
     \caption{Plates of Andromeda galaxy from the Asiago archive inspected in
     the M31-RV search.  The last two columns detail the number of plates
     obtained in 1988, when the outburst of M31-RV occurred, and in 1967,
     when an alleged previous outburst should have taken place.}
     \centering
     \begin{tabular}{ccrrrrrcc} 
     \hline
     &&\\
     \multicolumn{2}{c}{telescope}&
     \multicolumn{1}{c}{focal} &
     \multicolumn{1}{c}{first} &
     \multicolumn{1}{c}{last}  & 
     \multicolumn{1}{c}{N$_{tot}$}&
     \multicolumn{1}{c}{main} &
     \multicolumn{1}{c}{plates in} &
     \multicolumn{1}{c}{plates in} \\
     &&\multicolumn{1}{c}{length}&\multicolumn{1}{c}{plate}&\multicolumn{1}{c}{plate}
     &\multicolumn{1}{c}{plates}&\multicolumn{1}{c}{band}
     &\multicolumn{1}{c}{1967}&\multicolumn{1}{c}{1988}\\
     &&\\
     \hline
     &&\\
     1.22 m   & Newton     & 6.0 m & 29 Oct 1942 & 26 Aug 1992 & 831 & 795 in $B$ band & 25 & 2 \\
     1.22 m   & Cassegrain &19.1 m & 31 Oct 1961 & 27 Nov 1972 & 94  &  93 in $B$ band & 4  &   \\
     1.82 m   & Cassegrain &16.4 m & 05 Aug 1973 & 09 Dec 1988 & 194 & 177 in $B$ band &    & 2 \\
     67/92 cm & Schmidt    & 2.2 m & 02 Oct 1965 & 17 Dec 1993 & 291 & 264 in $B$ band & 2  & 8 \\
     40/50 cm & Schmidt    & 1.0 m & 14 Oct 1958 & 05 Mar 1986 &  37 &  28 in $B$ band &    &   \\  
     &&\\
     \hline
     \end{tabular}
     \end{table*}

Common to all models above is the uniqueness of the event: the progenitor
can experience a single such outburst in its life. In the Soker and Tylenda
(2003) and Retter and Marom (2003) approaches, it is the result of a merger
event that obviously cannot be repeated. In the Iben and Tutukov (1992) model an
extremely long time (a sizable fraction of a Hubble time) is required to
accrete at very low rates $10^{-2}$~M$_\odot$ on a WD that had to cool to
low temperatures.

     \begin{figure}
     \centerline{\psfig{file=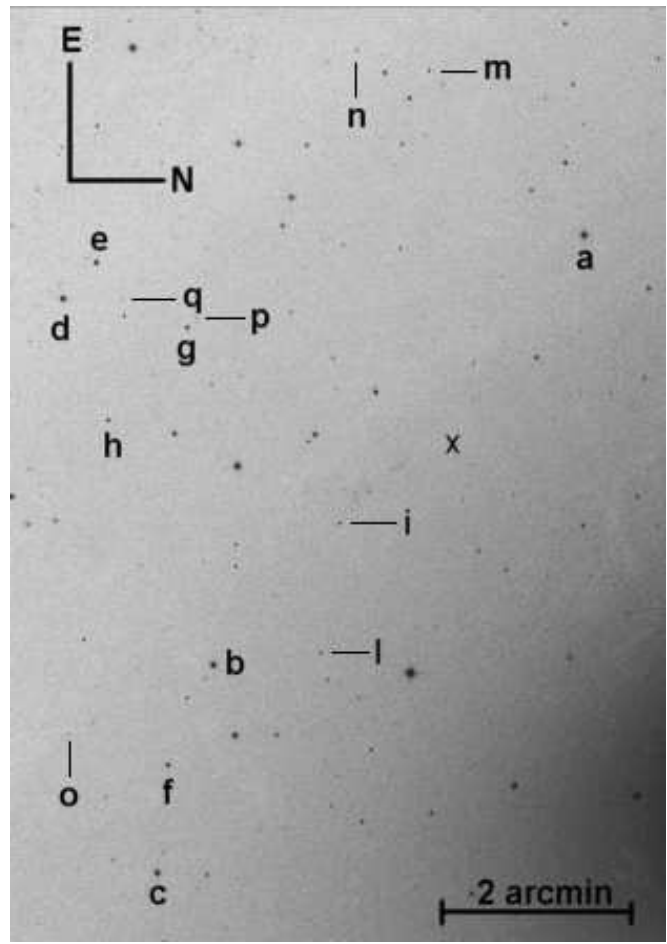,width=8.7cm}}
     \caption[]{Finding chart for the {\em BVR$_{\rm C}$I$_{\rm C}$} comparison 
     sequence listed in Table~2. The ``{\em X}'' marks the location of M31-RV.}
     \end{figure}

Therefore, the report by Sharov (1990) about a second outburst of M31-RV in
1967, 20 years before the main one, is something that deserves careful
scrutiny and independent verification. If confirmed, it would have profound
consequences on the theoretical modeling of M31-RV, V4332~Sgr and V838~Mon,
perhaps even more than the discovery of a massive and young B3~V companion
to the latter (Munari et al. 2002b).  The recent eruption of V838~Mon has
considerably revitalized the interest on this class of objects. In
anticipation of a growing modeling effort by the community, 
we decided to take advantage of the Asiago plate archive to
evaluate the reality of a second outburst of M31-RV and to investigate its
long term photometric evolution.

\section{Plate archive data}

Four instruments have contributed to the large collection of photographic
plates of M31 that we have located in the Asiago archive: the 1.22~m
and 1.82~m reflectors, and the 40/50~cm and 67/92~cm Schmidt telescopes.
Details about the number of plates, time span, focal length, limiting magnitude,
etc. are provided in Table~1.

In total, we have selected and retrieved 1447 plates of M31 from the Asiago
archive. They all have been inspected visually with a high quality Zeiss
binocular microscope. All plates have been inspected by the same author, and
about 10\% of them, randomly selected, checked by the other one. All key
plates have been inspected by both authors more than once (over a one month
time span and each time with a different orientation), taking care to make
them unrecognizable so to avoid biasing from memory of previous
inspections. The agreement between the estimates of the two authors and
their repeatability at different times turned out to be excellent, typically at
the 0.05 mag level, and very rarely differing by more than 0.1 mag.  Munari et
al. (2003) have determined an accurate astrometric position for M31-RV and
have provided a finding chart from one of the Asiago plates taken during the
1988 outburst.

A proper comparison sequence had to be established. We looked for literature
data of stars close to M31-RV and projected on similar background brightness
(thus roughly aligned parallel to local isophotes of the the unresolved
bulge of M31), to minimize biasing by the galaxy background when going from
plates of one instrument to those of another, taken with different focal
lengths, photometric bands, exposure times, seeing and sky conditions.  We
selected magnitudes obtained by Magnier et al. (1992) with CCD observations
that allowed proper handling of bulge background. The sequence we have
adopted is presented in Figure~1 and listed in Table~2.  Comparison with
other datasets (A.~Henden 2003, priv. comm.) indicate that there may be
errors in the 0.1~mag range for some of the stars, but using the ensemble
results in photometry close to the standard system.

     \begin{table}[!t]
     \caption{The comparison sequence plotted in Figure~1. Magnitudes from 
     CCD observations of Magnier et al. (1992).}
     \centering
     \begin{tabular}{crrrrll} 
     \hline
     &&\\
     &
     \multicolumn{1}{c}{$B$}&
     \multicolumn{1}{c}{$V$}&
     \multicolumn{1}{c}{$R_C$}&
     \multicolumn{1}{c}{$I_C$}\\
     &&\\
     \hline
     &&\\
     a &14.306      &13.691      &13.328      &13.058 \\
     b &14.780      &14.227      &13.902      &13.638 \\
     c &15.039      &14.695      &14.463      &14.269 \\
     d &15.272      &14.803      &14.440      &14.213 \\
     e &16.086      &15.217      &14.657      &14.275 \\
     f &16.644      &15.881      &15.487      &15.233 \\
     g &16.892      &16.099      &15.582      &15.239 \\
     h &17.401      &16.665      &16.184      &15.868 \\
     i &17.775      &17.154      &16.820      &16.662 \\
     l &18.369      &17.524      &17.051      &16.840 \\
     m &18.556      &18.116      &17.792      &17.472 \\
     n &19.043      &18.176      &17.649      &17.275 \\
     o &19.591      &19.589      &19.368      &------ \\
     p &20.117      &19.419      &18.562      &17.872 \\
     q &20.363      &19.729      &19.308      &18.981 \\
     &&\\
     \hline
     \end{tabular}
     \end{table}

The date, UT, telescope, filter, emulsion, exposure time and limiting magnitude
in the appropriate band for each one of the 1447 inspected plates is given
in Table~3 (available only in electronic form).

     \setcounter{table}{3}
     \begin{table}
     \caption{Results of the inspection of 1967 and 1988 plates looking for
     M31-RV.  For 1967, to save table length we list only sample plates for
     non-redundant dates (the whole list is accessible via electronic
     Table~3).  Additional plates for late 1966 and early 1968 are reported
     for completeness.}
     \centerline{\psfig{file=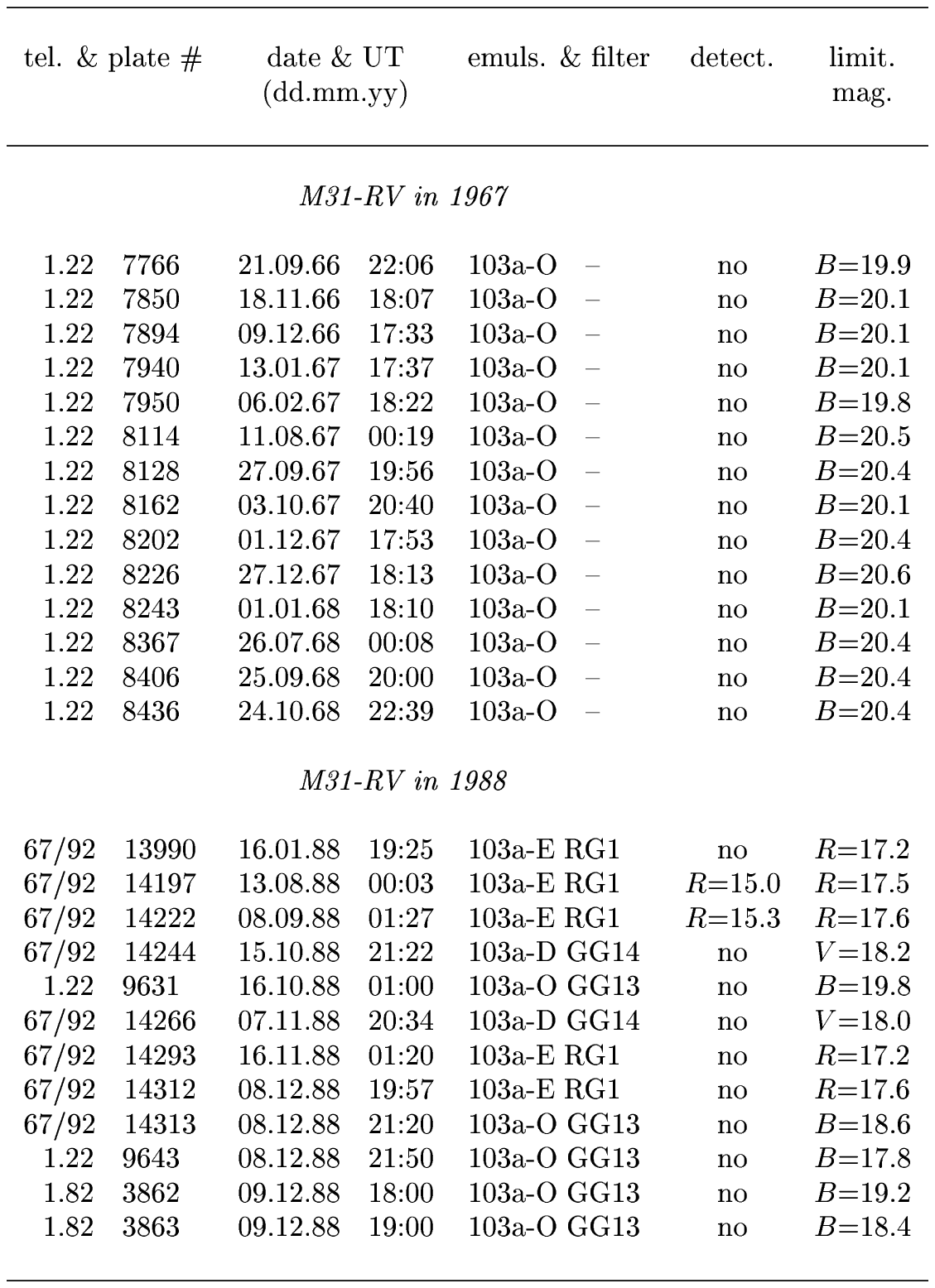,width=8.9cm}}
     \end{table}

\section{No outburst in 1967}

Sharov (1990) announced that M31-RV had twenty years earlier experienced
an outburst similar to that of 1988. He reported that while
inspecting a long series of plates of M31 taken with telescopes of the
Crimean Astrophysical Observatory he noted M31-RV around $B \sim$18.5 on
three plates taken on Aug 4, Sep 3 and Sep 4, 1967
(Sharov reported that the outburst occurred in 1968, but the JDs 
he tabulated leave no doubt it was 1967. The listed JDs 
firmly establish the 50~cm Maksutof telescope as the source instrument.
According to A.Tatarnikova (private communication) the focal length of this
instrument is 2.0~m).

The Andromeda galaxy has been frequently observed by Asiago telescopes for
half a century, mainly to search for novae within a long term program lead
by late Leonida Rosino (cf. Rosino 1973). More than 30 plates of M31 were
collected in 1967, and similarly in adjacent years.

Particularly useful are the plates taken at the Newton focus of the 1.22~m
telescope. It has a much larger aperture and longer focal length (6~m) than
the Sharov's 50~cm Maksutof camera, and its plates routinely show stars fainter
than $B$=20~mag close to the position of M31-RV (the limit away from the
bright background of the bulge of the Andromeda galaxy is generally one
magnitude fainter).

     \begin{table*}
     \caption{Compilation of all available optical and IR photometry of the
     1988 outburst of M31-RV. Only direct measurements are retained, derived
     or inferred ones being ignored.}
     \centerline{\psfig{file=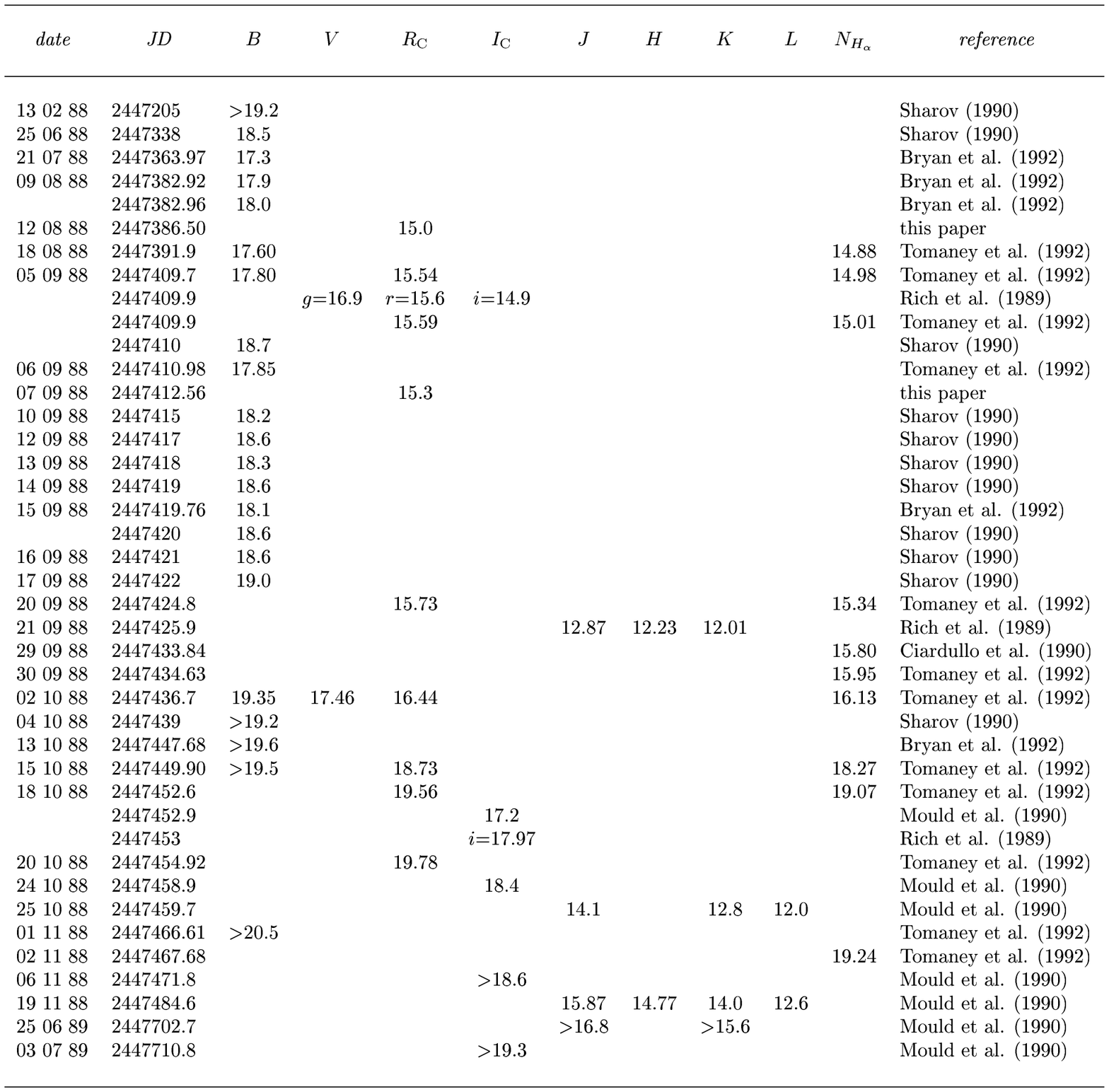,width=16.0cm}}
     \end{table*}

The 1.22~m plates rule out the outburst of M31-RV in 1967 announced by
Sharov (1990). Particularly useful is a plate taken on Aug~11 (cf.
Table~4), when M31-RV should have been at $B\sim$18.7 according to Sharov.
Nothing is present at the M31-RV position down to the local plate limit of
$B$=20.5. According to the 50 years covered by Asiago plates, the only
recorded event is that of 1988. It is worth noting that Goranskii et al. (2002)
have inspected archive plates spanning the time interval 1949--1994 in search
for previous outburst of V838~Mon, and found none.

The region of the bulge where M31-RV appeared is characterized by subtle dust
lanes and a knotty surface brightness distribution. It is possible that the
Crimean 50~cm Maksutof camera had trouble resolving the local 
inhomogeneities of the bulge brightness distribution, which could have been
confused for M31-RV on the 1967 plates. Alternative possibilities, such as a
gravitational lensing event or the appearance of a nova close to the
position of M31-RV do not apply because they should have been easily
visible on the deeper Asiago plates. It is worth noting that the Asiago 1.22~m
telescope discovered a sizable fraction of all novae
cataloged in M31 during the 1960's.

\section{The 1988 outburst}

\subsection{Evolution}

Table~5 collects all direct photometric observations that we have been able to
locate in literature concerning the 1988 outburst of M31-RV. Two $R_{\rm C}$
entries (cf. Table~3) come from the present inspection of plates from the
Asiago archive. By far, the best covered photometric bands are $B$, $R_{\rm
C}$ and $N_{\rm H_\alpha}$. The latter has been obtained with a narrow
filter centered on H$\alpha$ and characterized by a full width at half maximum (FWHM) of 75~\AA. Their
light-curves are presented in Figure~2. The $R_{\rm C}$ and $N_{\rm
H_\alpha}$ match well because the H$\alpha$ displayed a modest emission,
with negligible effect on the total flux through both $R_{\rm C}$ and
$N_{\rm H_\alpha}$ filters. More relevant is instead the fact that the
position of H$\alpha$ and therefore of the $N_{\rm H_\alpha}$ filter is
centered on the continuum that in M stars tries to emerge between the 6200
and 6700~\AA\ TiO bands.  $N_{\rm H_\alpha}$ tends to appear brighter
compared to $R_{\rm C}$ as the spectral type progresses from M0 to M5. At
later spectral types the two bands converge back to similar values because
the rapidly increasing steepness of the spectrum increases the flux in the red
wing of the $R_{\rm C}$ band. The small differences in Figure~2 between the
$R_{\rm C}$ and $N_{\rm H_\alpha}$ branches therefore seem to just reflect
the monotonic evolution toward later M spectral types of the M31-RV
continuum.

     \setcounter{figure}{1}
     \begin{figure*}
     \centerline{\psfig{file=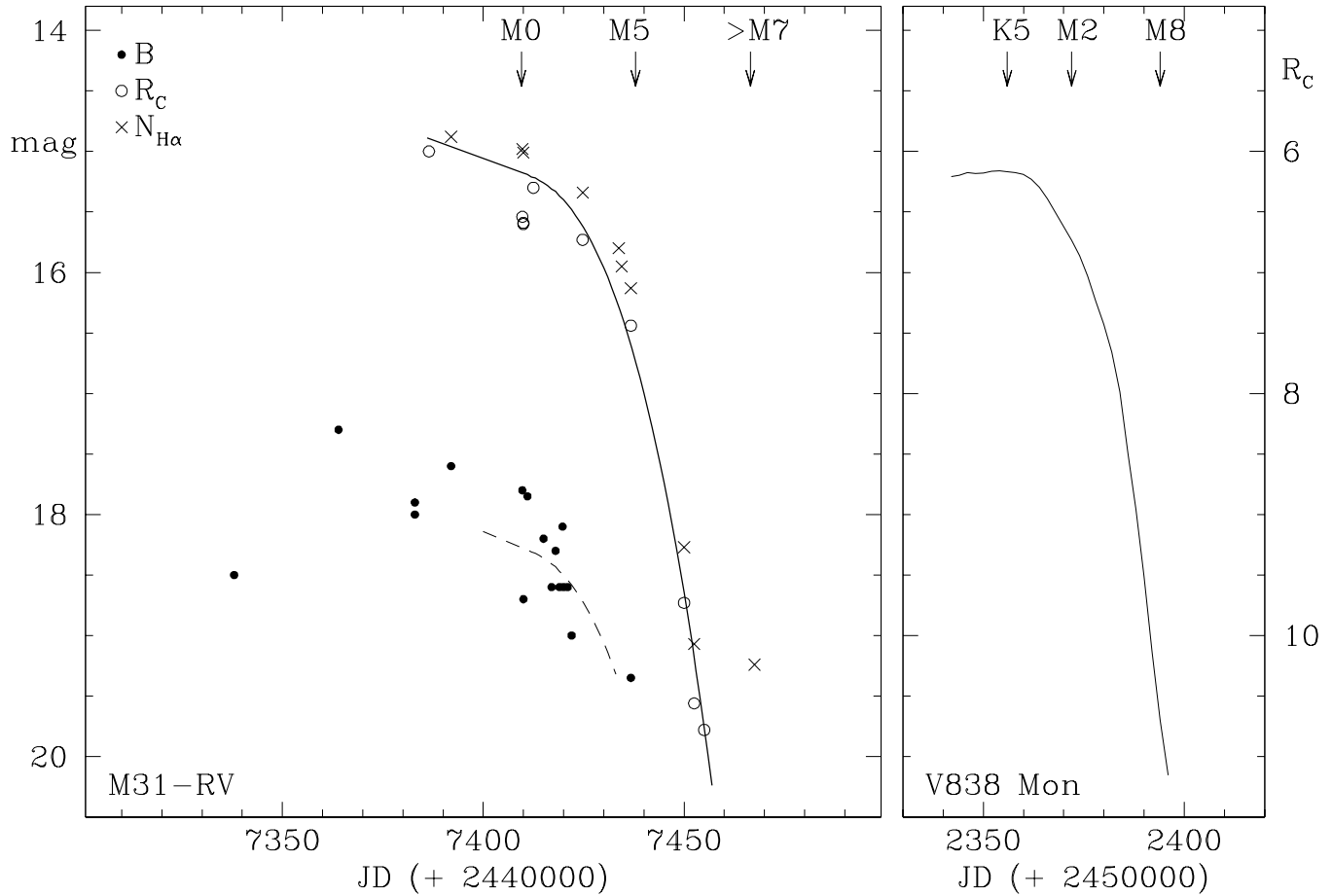,width=18.0cm,angle=270}}
     \caption[]{Photometric and spectroscopic evolution of M31-RV (spectral
     classifications from Rich et al. 1989 and Mould et al. 1990). The 
     corresponding section of the $R_{\rm C}$ lightcurve of V838~Mon
     is shown for comparison on the same scale (spectroscopic classification 
     from Munari et al. 2002c).}
     \end{figure*}

The $R_{\rm C}$ lightcurve of M31-RV is less scattered compared to the $B$ lightcurve
for a number of reasons. The literature $R_{\rm C}$ data come mainly from
CCD observations, and the two Asiago $R_{\rm C}$ data-points are relative to a
comparison sequence calibrated via CCD observations. The comparison
sequences used in the literature to derive $B$ data are of unknown
origin. Furthermore, the $R_{\rm C}$ data are obtained with accurate detector +
filter pairs well matching the standard system, while several of the $B$ band
data-points are not color corrected or come from scattered emulsion+filter
combinations (which are relevant in the case of the very red colors
displayed by M31-RV). Finally, the contrast between M31-RV and the
background bulge brightness was more favorable in $R_{\rm C}$ than in $B$.

The striking similarity of the $R_{\rm C}$ light-curves of M31-RV and
V838~Mon is evident in Figure~2 (V838~Mon $R_{\rm C}$ data are taken from
Munari et al. 2002c and Bond et al. 2003). The comparison is obviously
limited to the portion of the lightcurve of M31-RV covered by the
observations (while the V838~Mon one extends well beyond the small
displayed section). Both objects, after a plateau phase characterized by a
slowly evolving K-type spectrum, experienced a sudden drop of several
magnitudes (reaching $\Delta R_{\rm C}$=0.2~mag~day$^{-1}$ for M31-RV and
$\Delta R_{\rm C}$=0.3~mag~day$^{-1}$ for V838~Mon) accompanied by a
corresponding temperature drop as indicated by the spectral type sweeping quickly
through the M-type sequence toward classifications so far seen only in brown
{\em dwarfs} (cf. Evans et al. 2003). As evident from the evolution of
reddening and spectral energy distribution discussed in following sections,
the drop in magnitude of M31-RV is not due to dust condensation in the
ejecta, but instead mainly due to drop in temperature during the expansion
(shifting progressively the emission peak toward the IR) and to an overall
decrease in luminosity.

\subsection{Reddening}

From available data it is possible to estimate at different epochs 
the reddening affecting M31-RV.

At the time of the {\em JHK} observation from of Sep 21, 1988 reported in
Table~5, the spectral type of M31-RV was close to M2 (cf. Figure~2). 
According to Frogel and Whitford (1987), the intrinsic color of M2 giants in
the Bulge of our Galaxy (taken to resemble their counterparts in the Bulge
of M31, with comparable ages and metallicities, cf. Davidge
2001) is ($J-K$)$_\circ$=0.81.  Compared with the observed $J-K$=0.86 for
M31-RV, it implies $E_{J-K}$=0.05. The relation between $E_{J-K}$ and
$E_{B-V}$ for M2 giants in the KPNO infrared system is (Fiorucci and Munari
2003):
\begin{equation}
\frac{E_{J-K}}{E_{B-V}}=0.596 + 0.005\times E_{B-V}
\end{equation}
and the corresponding reddening toward M31-RV is therefore $E_{B-V}$=0.08. At 
the time of the infrared observations of Oct 25, 1988, the spectral type was
$\sim$M7, for which ($J-K$)$_\circ$=1.23 (again from Frogel and Whitford
1987). Compared with the observed $J-K$=1.30 it gives $E_{J-K}$=0.07 and
correspondingly $E_{B-V}$=0.12. The latest IR observation in Table~5 cannot
be used because the spectral classification at that time is unknown (by
analogy with V838 Mon it was probably later than M10).

By the time M31-RV was passing through the shoulder of the $R_{\rm C}$
lightcurve in Figure~2 (JD$\sim$2447418), the optical color was $B-R_{\rm
C}\approx$+3.1 and the spectral type $\sim$M1. From Kurucz models computed
on purpose for the M31 bulge metallicity ([Fe/H]=$-$0.2), the intrinsic
color of an M1 supergiant is $B-R_{\rm C}$=2.79. The excess is therefore
$E_{B-R}$=0.31. The transformation relation between $E_{B-R}$ and $E_{B-V}$ (both in
the Landolt realization of the Johnson and Cousins systems) for early M
giants and a normal extinction law ($R_V = A_V / E_{B-V} = 3.1$) is (from
Fiorucci and Munari 2003):
\begin{equation}
\frac{E_{B-R}}{E_{B-V}}=2.044 + 0.099\times E_{B-V}
\end{equation}
This gives $E_{B-V}$=0.15 for M31-RV. 

The three independent determinations consistently converge toward:
\begin{equation}
E_{B-V}=0.12 ~\pm0.02
\end{equation}
as the reddening affecting M31-RV, with no indication of any significant
increase during the abrupt photometric descent from optical maximum
brightness of M31-RV, which cannot therefore be ascribed to dust
condensation in the ejecta. A sizable fraction of the total reddening
affecting M31-RV arises in our own Galaxy. In fact, the Burnstein and Hales (1982)
extinction maps report $E_{B-V}\sim$0.1 as the total Galactic extinction along
the line of sight to M31.

\subsection{Absolute magnitude}

The distance modulus to M31 has been recently determined as 
24.49$\pm$0.11~mag by Joshi et al. (2003) and as 24.47$\pm$0.08~mag by
Stanek and Garnavich (1998). Taking the average of 24.48~mag and the
$E_{B-V}=0.12$ reddening from the previous section, the absolute magnitude
of M31-RV at peak $R_{\rm C}$ brightness around Aug 15, 1988 ($R_{\rm
C}\sim$14.94) is $M_{R_C}\sim -$9.88. The true maximum could have been even
brighter because the lightcurve is not completely mapped. Repeating the
exercise for the $B$ band, the maximum can be estimated to have occurred
around JD 2447362 at $B$=17.4, to which it corresponds $M_B$=$-$7.7.

The bolometric correction and $V-R_{\rm C}$ for M0 supergiants are $-$1.29 and
+0.97, respectively (Drilling and Landolt 2000). According to Figure~2,
M31-RV was at $R_{\rm C}\sim$15.2 by the time it was classified M0 by Rich
et al. (1989), which implies $M_{bol}\sim -$9.95 and $L\sim
7.5\times10^5$~L$_\odot$, i.e. one of the brightest stars in M31 and the
whole Local Group.

\subsection{An universal eruption mechanism ?}

The striking photometric and spectroscopic similarities between M31-RV and
V838~Mon suggest a similar outburst mechanism. The absolute magnitude 
reached by the two events also seems quite similar.

Figure~2 indicates that both M31-RV and V838~Mon when transitioning from the
plateau to the rapid fading phase were displaying a $\sim$M1 supergiant
spectrum. At that time, the absolute magnitude of M31-RV was $M_{R_{\rm C}}
\approx -$9.6.  At the corresponding time the magnitude of V838~Mon was
$R_{\rm C}$=6.2. The reddening affecting V838~Mon is uncertain, but a fair
estimate is $E_{B-V}$=0.5 (cf. Munari et al. 2002a). Assuming the same
absolute magnitude of M31-RV, this corresponds to a distance $d_{\rm
V838~Mon}=8$~kpc. This value is in good agreement with the average of the
distance determinations by Bond et al. (2003) based on the HST imaging of
the V838~Mon light echo, and Munari et al. (2002b) spectrophotometric
distance to the B3~V component in the V838~Mon binary. Therefore 
the photometric and spectroscopic evolution of V838~Mon and M31-RV were
similar, as well as the absolute magnitude at the time the drop in
temperature occurred.

Such similarities are remarkable in view of the different ages and evolution
histories of the two objects. M31-RV appeared in the Bulge of M31, which is
characterized by a turn-off mass around 1~M$_\odot$ and a high metallicity
[Fe/H]=$-$0.2. V838~Mon appears instead to be young and massive (the
companion to the erupted component is a B3~V star) and it is located in the
outskirts of the galactic disk, at galacto-centric distances of 15-17~kpc,
where the metallicity is lower and of the order of [Fe/H]=$-$0.6 (cf.
Davidge 2001). Yet, the two events show the same evolution and absolute
luminosity in $R_{\rm C}$. This seems to suggest that an {\em universal}
explosion mechanism could have powered both events, a mechanism independent
from the way in which a stellar system reaches it. The independence of the
outcome from the initial conditions is a characteristic, for example, of models 
of SN~Ia, well known for their homogeneity in absolute magnitude and lightcurve
shapes. In SN~Ia, a WD reaches the Chandrasekhar mass and ignites carbon
burning, irrespective of whether a merger of two WDs or the
accretion on a single WD from a non-degenerate companion occurred. Here we postulate
that a common eruption mechanism must have powered both M31-RV and V838~Mon,
the outcome of which was not affected by the large differences in metallicity, age
and mass of the two objects.

The theoretical models so far published do not seem able to explain both
M31-RV and V838~Mon, as well as the similarity of the two events. The Iben
and Tutukov (1992) TNR mechanism cannot work
in V838~Mon, because the young age implied by the presence of a B3V star in
the system is too short for a WD to cool and accrete enough material at a
very low accretion rate. Both Soaker and Tylenda (2003) and Retter and Marom
(2003) suggestions of swallowed stellar or planetary companions by an
expanding giant seem unable to account for the strong similarities exhibited
by M31-RV and V838~Mon. The results presented in this paper therefore
support the need of a radically new model if M31-RV, V838~Mon and V4332~Sgr
are to be explained as a {\em homogeneous class} of astronomical objects.

\acknowledgements{We would like to thank Arne Henden for careful reading and
commenting the manuscript before submission. Thanks also go to Maria Antonia
Rossi for assistance in checking for typos in the tabulated electronic data,
and to Sergio Dalle Ave and Alfredo Segafredo for Figure~1 preparation.}

\end{document}